\documentclass[epj]{svjour}
\usepackage{graphics}
\usepackage{times}

\newcommand{\llangle}{\langle\kern-0.23em\langle}
\newcommand{\rrangle}{\rangle\kern-0.23em\rangle}

\begin{document}

\sloppy

\title{Magnetostatic field noise near metallic surfaces}

\author{Carsten Henkel}

\institute{%
Institut f\"{u}r Physik, Universit\"{a}t Potsdam,
Am Neuen Palais 10, 14469 Potsdam, Germany
\and
Laboratoire Charles Fabry de l'Institut d'Optique, B.P. 147,
91403 Orsay cedex, France
}%
\date{rec 25 Mar 2005 / rev 17 May 2005}
%
\abstract{
We develop an effective low-frequency theory
of the electromagnetic field in equilibrium with thermal objects. 
The aim is to compute 
thermal magnetic noise spectra close to metallic
microstructures. We focus on the limit where the
material response is characterized by the 
electric conductivity. At the boundary
between empty space and metallic microstructures, 
a large jump occurs in the dielectric function which leads to a 
partial screening of low-frequency magnetic fields generated by thermal 
current fluctuations. We resolve a discrepancy between two approaches 
used in the past to compute magnetic field noise spectra close to 
microstructured materials.
} 

\maketitle
\section{Introduction}
\label{intro}

In the context of atom chips, low-frequency thermal magnetic noise 
has recently emerged as one crucial element that limits the lifetime 
of miniaturized atom traps. Recent experiments 
\cite{Cornell03a,Jones03,Vuletic03}
have confirmed the basic features of theoretical predictions for spin 
flip processes induced by magnetic near field fluctuations. 
A current trend is 
to extract other physical mechanisms that lead to loss
by subtracting the near-field induced loss rate.
One particularly interesting mechanism is the lowering of the trap
depth due to atom-surface interactions \cite{Vuletic03}. 
Accurate calculations of magnetic near field noise are clearly needed 
for this purpose.
Magnetic fluctuations are also relevant in other contexts, for example in 
biophysics where they impose ultimate limits on the sensitivity of SQUID 
detectors \cite{Varpula84}, and in
magnetic resonance force microscopy, a near-field variant of magnetic 
resonance imaging \cite{Sidles95,Sidles00}.

Typically, one is interested in field frequencies $\hbar\omega \ll k_{B} 
T$ where the noise is dominantly classical. The border to the 
quantum regime can be reached with magnetically trapped atoms,
either by cooling the microtrap components and/or applying strong 
static magnetic fields that push up the relevant frequency range
(given by the Larmor frequency). 
Even at the highest frequencies conceivable with state-of-the-art 
atom chip structures (in the GHz range), the (vacuum) field wavelength 
$\lambda$ is much larger than the characteristic distances, so that 
the quasistatic approximation applies outside the structures. This 
leads to a peculiar situation to describe the field fluctuations. Indeed,
one cannot apply the standard procedure and attribute thermal or 
quantum fluctuations to the normal mode amplitudes of the field, 
because there are no nontrivial solutions to the 
homogeneous field equations (i.e., eigenmodes) in the quasistatic limit. 
Near field 
noise is actually dominated by the fluctuations of its sources (currents,
magnetic moments) whose spectral mode density depends on material 
or atomic constants \cite{Varpula84,Barnes98,Pendry99b,Scheel00b,Savasta00}.
As a result, the near field noise spectrum 
differs markedly from the celebrated blackbody radiation law 
\cite{Varpula84,Shchegrov00,Joulain03}.

Roughly, two approaches can be identified to compute magnetic noise 
close to micro- and nanostructures. The first one can be traced back 
to the fluctuation electrodynamics put forward by Rytov and co-workers
\cite{Rytov3} in the 1950's. Based on a statistical thermodynamics 
argument (the fluctuation-dissipation -- FD -- theorem \cite{Callen51}), 
random charge and current fluctuations are associated to
a dissipative material structure. Their radiation is incoherently summed 
to give the total noise strength of the field. In a planar geometry, 
the radiated field and the required averaging can be 
calculated analytically. Results along this line have been computed 
and experimentally verified for planar metallic layers by Varpula and Poutanen 
in 1984 \cite{Varpula84}. Sidles and co-workers give an extensive 
discussion with applications for magnetic resonance microscopy and
quantum computing~\cite{Sidles00}.

An alternative approach uses the FD theorem 
for the (magnetic) field itself and has been popularized in a series 
of papers by Agarwal in 1975 \cite{Agarwal75a}. The advantage is that 
the incoherent averaging is avoided; the FD theorem reduces the calculation 
to the radiation of a single dipole source (Green function), 
located at the observation point. This method has been applied, in the 
context of atomic microtraps, by the 
present author and co-workers \cite{Henkel99c} and Rekdal and 
co-workers \cite{Scheel04a,Scheel05a}.
Both approaches have been shown to be equivalent under fairly general 
conditions, thanks to an identity that implements
the FD theorem in electromagnetism \cite{Eckhardt82,Henry96}.

In the context of integrated atom optics, incoherent summation over 
fields has been put forward by S. P\"{o}tting and the present author 
as a versatile tool to handle arbitrary nanostructures~\cite{Henkel01a}. 
It just remains to perform a certain spatial integral over the volume 
filled with electrically conducting material. This yields the correct 
scaling of the noise spectrum with the atom chip geometry, provided
the skin depth is long enough.
Based on this approach,
for example, the Vuletic group could fit reasonably well spin flip loss 
rates close to rectangular wires \cite{Vuletic03}.  (See also
Refs.~\cite{Scheel05a,Dikovsky05} for a re-analysis and discussion.)
A closer comparison shows, however, that the theoretical results are 
off by numerical 
factors between two and three compared to the noise spectrum 
predicted by the FD theorem~\cite{Henkel01a}. This discrepancy is the 
motivation for the present paper.  We point out an error in the 
`incoherent summation' approach that is linked to the particular 
boundary conditions for the electromagnetic field at the surface of a 
good conductor. We derive approximate boundary conditions 
that apply to any geometry in the low-frequency range relevant for
atom chips.  In the planar case, we show that they
lead to an accurate agreement with the `Green function' approach in
the limit that the vacuum wavelength is the largest length scale. The 
only point missing in the theory is the blackbody noise level that 
prevails at large distances, but this one is in most situations impossible 
to detect anyway. 

\begin{figure}[bth]
\begin{center}
    \resizebox{105mm}{!}{%
\includegraphics*{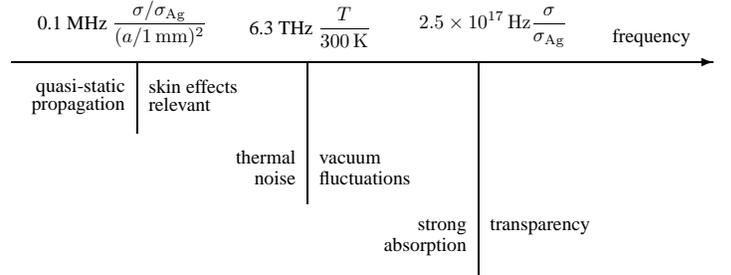}
    }
\end{center}
\caption[]{Characteristic frequencies involved in near field 
electromagnetic noise. The conductivity of silver, $\sigma_{\rm Ag}$ 
is used as a convenient scaling parameter. A typical geometrical 
feature size is denoted by $a$. Three characteristic frequencies 
separate different regimes as illustrated by the vertical lines.
The formulas at the frequency axis give explicit values and their
scaling with the relevant parameters. 

For example, on a scale
$a = 10\,\mu{\rm m}$, the skin effect is irrelevant for the field 
propagation near a silver structure at frequencies below $0.1\,{\rm MHz}
\times 10^4 = 10^3\,{\rm MHz}$; the quasi-static approximation applies in 
this regime. At frequencies approaching $2.5\,10^{17}\,{\rm Hz}$, silver
becomes transparent, and its permittivity does no longer involve
a purely real conductivity.}
\label{fig:regimes}
\end{figure}

The parameter regime we focus on in this paper is illustrated in 
Figure~\ref{fig:regimes}. We shall call `quasi-static' the regime 
where the skin depth $\delta$ inside the material is larger than any other 
geometrical scale (denoted $a$ in Fig.\ref{fig:regimes}). We focus on
metallic materials and use the 
definition $\delta = (\mu_{0} \sigma \omega / 2)^{-1/2}$ in terms of 
the (DC) conductivity $\sigma$. Our theory
aims at covering both the quasi-static regime and a skin depth comparable 
to $a$. The 
temperature $T$ defines another frequency scale below which the field 
fluctuations behave classically. This is actually not a limitation 
as long as we assume thermal equilibrium. The theory is 
extended into the quantum regime 
with the replacement $k_{B}T \mapsto 
\frac12 \hbar\omega \coth( \hbar\omega / 2k_{B}T)$ 
and assuming symmetrized noise correlation spectra. 
At frequencies in the visible and ultraviolet range, however, the 
dielectric function of the material becomes complicated, and 
more parameters (transverse optical phonon frequency,
plasma frequency, tabulated data \ldots) are needed for an accurate 
modelling. 

The present paper thus aims at clarifying the validity of the 
`incoherent summation' approach and at extending it into the regime 
of a short skin depth using the appropriate boundary conditions. We 
also argue that from a practical point of view, the FD approach 
appears simpler because it is sufficient to compute the radiation from 
a single point source, while for `incoherent summation', many sources 
(anywhere inside the spatial domains filled with absorbing material) 
have to treated. 

In the following sections, we start by writing down the basic equations 
for the magnetic field and explain the relevant parameter regimes
(Sec.\ref{s:jumps}). The boundary condition at the surface of a good 
conductor is derived. In section~\ref{s:summation}, we review the 
incoherent summation technique and show for the special case of a 
metallic half-space that with the correct boundary condition, one gets 
a magnetic noise spectrum in agreement with the FD approach. We show 
that in the limit of a short skin depth, the transmission of the 
magnetic field out of the metal becomes much less efficient. In 
subsection~\ref{s:layer-discussion}, we give a qualitative 
explanation of the power laws in the distance-dependence of the noise
spectrum and review results obtained for a 
thin metallic layer. In section~\ref{s:FD-technique}, we formulate 
the equations to be solved within the FD approach when the 
quasistatic approximation is made in the spatial domains filled with 
vacuum. The formulation applies to an arbitrary geometry and is then 
specialized to a metallic half-space. In the latter case, we 
demonstrate 
agreement with the more complex, fully retarded FD approach in the 
long vacuum wavelength limit.

\section{Boundary conditions at low frequency}
\label{s:jumps}

We want to solve the Maxwell equations in a non-magnetic medium 
($\mu( {\bf x} ) \equiv \mu_{0}$):
\begin{eqnarray}
\nabla \cdot ( \sigma {\bf E} + {\bf j} ) = 0,
&&
\quad
\nabla \cdot {\bf B} = 0,
\label{eq:Maxwell-E}
\\
\nabla \times {\bf E} = {\rm i} \omega {\bf B},
&&
\quad
\nabla \times {\bf B} = \mu_0
\left( 
\sigma {\bf E} 
+ {\bf j} 
+ \nabla \times {\bf M}
\right).
\label{eq:Maxwell-B}
\end{eqnarray}
where $\sigma( {\bf x})$ is the metal conductivity.  In the vacuum
regions, $\sigma ( {\bf x}) = - {\rm i} \varepsilon_{0} \omega$ which
will be assumed much smaller in magnitude than the metal conductivity,
denoted $\sigma$.  Two kinds of sources appear here that apply to
the methods of magnetic noise calculations mentioned in the
introduction.  They are represented by the externally prescribed terms
${\bf j}( {\bf x} )$ and ${\bf M}( {\bf x} )$.  In the incoherent
summation technique, the current density ${\bf j}( {\bf x} )$ is
localized inside the metal and represents thermal fluctuations.  It is
a random quantity with correlation function given by
Eq.(\ref{eq:current-FD}).  For the calculation of the magnetic Green
function, the magnetization ${\bf M}( {\bf x} )$ can be identified
with the magnetic moment of an atom localized in vacuum outside the
metal, see after Eq.(\ref{eq:div-A-0}).  Depending on the method of
calculation, only one or the other source term is actually nonzero.

Taking the curl of~(\ref{eq:Maxwell-B}b), 
we find the following wave equation for the magnetic field
\begin{eqnarray}
\nabla^2 {\bf B} 
+ q^2 {\bf B} 
& = &
-\mu_0 \left[
\nabla \times {\bf j}( {\bf x} )
- \nabla \times \nabla \times {\bf M}
\right.
\nonumber
\\
&& \left. {}
+ \left( \nabla \sigma \right) \times {\bf E} 
\right].
\label{eq:B-wave}
\end{eqnarray}
Outside the metal, $q( {\bf x} ) = \omega/c$. Inside the metal,
$q( {\bf x} )$ is complex and related to the skin depth 
$\delta$ by
$q^2 = 2 {\rm i} / \delta^2 = {\rm i} \mu_{0} \sigma \omega$.
We could also introduce a spatial diffusion coefficient $D = \mu_0 \sigma$ 
by going back to time-dependent equations. 
The source terms of the magnetic wave equation~(\ref{eq:B-wave}) 
are related to the current density, the
magnetization density and to a current parallel to the surface 
induced by the electric field.
It turns out that this last
current leads to a jump in the derivative of the magnetic field 
across the interface. The solutions to the wave equation can be sought 
in terms of a Green function, assuming localized sources. 
This is why we assume in the following that the current ${\bf j}( {\bf 
x} )$ is nonzero only inside the metal.

\subsection*{Boundary conditions}

The magnetic field itself is continuous across the interface of the 
metal (assumed to be non-magnetic). The electric field components 
parallel to the interface are continuous as well. For the normal 
components, the continuity of the displacement field gives
\begin{equation}
    \frac{ 2 {\rm i} }{ \delta^2 } 
    \left. {\bf n} \cdot {\bf E} \right|_{\rm in} 
    = \frac{ (2 \pi)^2 }{ \lambda^2 }
    \left. {\bf n} \cdot {\bf E} \right|_{\rm out}
    \label{eq:normal-E}
\end{equation}
We shall assume that the wavelength $\lambda$ is much larger than any 
other relevant length scales in the problem, and take the limit
$\delta \ll \lambda \to \infty$. To lowest order, this transforms 
Eq.(\ref{eq:normal-E}) into
\begin{equation}
    \left. {\bf n} \cdot {\bf E} \right|_{\rm in} 
    = 0.
    \label{eq:normal-E-2}
\end{equation}
We check explicitly below that this boundary condition 
yields results consistent with a fully retarded calculation.

We next calculate the jump condition for the magnetic field due
to the surface current. Assume first a planar metallic surface located
at $z = 0$ with unit outward normal ${\bf n}$. Noting that
\(
\nabla \sigma( {\bf r} ) = - \sigma {\bf n} \delta( z )
\)
and integrating the magnetic wave equation
along a path perpendicular to the interface, 
we find the following jump condition
\begin{equation}
\left.
\frac{ \partial }{ \partial {n} } {\bf B}
\right|_{\rm in}^{\rm out} = \mu_0 \sigma {\bf n} \times {\bf E}
\label{eq:jump-B}
\end{equation}
where the scripts `in' and `out' mark the field inside and outside the
metal. The electric field is taken at the interface because only its 
tangential components are involved.

We can also get this boundary condition directly from the Maxwell 
equations and without specifying a planar boundary.
At the metal surface, we evaluate~(\ref{eq:Maxwell-B}b) just above and
below the interface and find, using that the current ${\bf j}$
vanishes at the interface,
\begin{equation}
\left.
\nabla \times {\bf B} 
\right|_{\rm in}^{\rm out}
=
\left.
\mu_0 \sigma {\bf E}
\right|_{\rm in}^{\rm out}
=
-
\left.
\mu_0 \sigma {\bf E}
\right|_{\rm in}
\end{equation}
The last equality is again due to the negligible vacuum conductivity.
(Note that $\varepsilon_{0} \omega / \sigma = (2\pi\delta / 
\lambda)^2/2 \to 0$ in lowest order.)
Taking components parallel to the interface, we find
\begin{equation}
\left.
{\bf n} \times ( \nabla \times {\bf B}) 
\right|_{\rm in}^{\rm out}
=
-
\mu_0 \sigma \,
{\bf n} \times {\bf E}
\label{eq:jump-B-2}
\end{equation}
Now, from $\nabla \cdot {\bf B} = 0 = \partial_k (\nabla \cdot {\bf B})
= \partial_i \partial_k B_i$, we can derive using the Gauss theorem:
\(
\left.
{n}_i \nabla {B}_i 
\right|_{\rm in}^{\rm out} = 0
.
\)
This identity cancels one of the terms coming from the expansion of 
the double vector product in~(\ref{eq:jump-B-2}). We thus find the
jump condition~(\ref{eq:jump-B}). Note that the present derivation
is valid for any geometry of the interface---which is less obvious for
the previous one because one has to integrate the Laplacian operator.

\section{Incoherent summation}
\label{s:summation}

In this section, we focus on a planar metallic surface and the 
quasi-static limit (geometrical distances even smaller than the skin depth). 
We display the surface electric field and the correction it implies 
for the transmitted
magnetic field. Technical details are deferred to 
Appendix~\ref{a:summation}.


\subsection{Transmitted field expansion}

For a planar surface parallel to the $xy$-plane, an expansion in plane 
waves with two-dimensional wave vectors is straightforward. We use
the notation ${\bf K} = (k_x, k_y, 0)$ and find just 
below the metal surface the following expansion
\begin{equation}
{\bf B}( {\bf r} ) 
= \int\!\frac{ {\rm d}^2 K }{ (2\pi)^2 }
\left(
{\bf B}_{i}[ {\bf K} ]
{\rm e}^{ {\rm i} {\bf k}_{i} \cdot {\bf r} }
+
{\bf B}_{r}[ {\bf K} ]
{\rm e}^{ {\rm i} {\bf k}_{r} \cdot {\bf r} }
\right)
\label{eq:2D-Fourier}
\end{equation}
where ${\bf k}_{i,r} = {\bf K} \pm {\rm i} \kappa {\bf n}$ with
$\kappa = \sqrt{ K^2 - q^2 }$ (${\rm Re}\, \kappa > 0$). 
In the quasi-static limit, we have
$K \gg |q|$, and therefore $\kappa \approx K$. More general formulas
can be found in Appendix~\ref{a:summation}.  
As shown there, a current density 
localized
below the metal surface produces an `incident' magnetic field with
Fourier transform
\begin{equation}
{\bf B}_{i}[ {\bf K} ] = \frac{ {\rm i} \mu_0 }{ 2 K } {\bf k}_{i} 
\times {\bf J}[ {\bf K} ]
\label{eq:Bi-1}
\end{equation}
where 
\begin{equation}
{\bf J}[ {\bf K} ]
= \int_{-\infty}^0\!{\rm d}x_3 \, {\rm e}^{ K x_3}
{\bf j}[ {\bf K}; x_3 ]
\label{eq:def-J-K}
\end{equation}
with ${\bf j}[ {\bf K}; x_3 ]$ being the 2D spatial Fourier transform of 
${\bf j}( {\bf x} )$. 
For the solution of the reflection/transmission problem, we also need 
the electric field at the interface. Its tangential components are given by
\begin{equation}
{\bf n} \times {\bf E}[ {\bf K} ]
=
- \frac{ {\bf n} \times {\bf K} }{ \sigma K }
( {\bf k}_{i} \cdot {\bf J}[ {\bf K} ] )
= 2 {\bf n} \times {\bf E}_{\rm i}[ {\bf K} ]
.
\label{eq:E-tang-solution}
\end{equation}
The factor of two between the `incident field' and the 
actual field at the interface, may be explained intuitively
by working with image currents (or dipoles): to fulfill the boundary
conditions, one combines the fields
of the actual dipole (inside the metal) and of an image dipole. Since the
field is incident from a region with a large `refractive index', the reflected 
field has the opposite sign: source and image dipoles therefore have the
same polarity if they are parallel to the interface.
As a consequence, their field components parallel to the interface double.
Dipoles perpendicular to the interface have mirror images with the
opposite polarity, leading to the cancellation of the normal field 
component, as required by the boundary condition~(\ref{eq:normal-E-2}).

Combining the jump condition~(\ref{eq:jump-B}) for the magnetic field
with the in-plane electric field~(\ref{eq:E-tang-solution}), we 
get the transmitted magnetic field:
\begin{equation}
{\bf B}_{\rm t}[ {\bf K} ] = 
\frac{ {\rm i} \mu_0 }{ 2 K }
{\bf k}_{t} \times {\bf J}[{\bf K}] 
+ 
\frac{ \mu_0 }{ 2 K^2 }
({\bf n}\times{\bf K}) ( {\bf k}_{t}\cdot{\bf J}[{\bf K}] )
\label{eq:Bt-solution}
\end{equation}
where ${\bf k}_{t} = {\bf K} + {\rm i} K {\bf n}$.
With respect to the incident field~(\ref{eq:Bi-1}), we thus have
an additional term with components parallel
to the interface. This term is absent when computing the field generated 
by the current density ${\bf j}$ as if the latter were located in vacuum.
It ensures in particular that the transmitted field vanishes if ${\bf 
j}$ is parallel to ${\bf n}$, which is a well-known result for a 
planar geometry (see, e.g., \cite{Varpula84}). We show now that 
the interference between both terms reduces the field fluctuations 
for some components, and reproduces the noise tensor found asymptotically 
from a retarded calculation.

\subsection{Magnetic field correlation tensor}

We first compute the correlation function of the Fourier transformed
current. Local thermodynamic equilibrium gives the basic 
relation~\cite{Rytov3,Scheel99a} for the current fluctuations
\begin{equation}
\left\langle
j^*_k( {\bf x}_1 ; \omega )
j_l( {\bf x}_2 ; \omega' )
\right\rangle
=
2\pi S( {\bf x}_{1}; \omega) \delta_{kl}
\delta( {\bf x}_1 - {\bf x}_2 )
\delta( \omega - \omega' )
,
\label{eq:current-FD}
\end{equation}
and we find
\begin{equation}
\left\langle
J^*_k[ {\bf K}_1 ; \omega ]
J_l[ {\bf K}_2 ; \omega']
\right\rangle
=
\frac{ (2\pi)^3 }{ 2 K_1 } 
S( \omega) \delta_{kl}
\delta( {\bf K}_1 - {\bf K}_2 )
\delta( \omega - \omega' )
\label{eq:JJ-correlation}
\end{equation}
The noise spectrum behaves like $S({\bf x}; \omega) \approx 2 \sigma( 
{\bf x} )  k_{\rm B} T$ 
at low (sub-thermal) frequencies. The magnetic noise tensor 
${\cal B}_{ij}( {\bf x}_1, {\bf x}_2 )$ is defined by
\begin{equation}
\left\langle
B^*_i( {\bf x}_1 ; \omega )
B_j( {\bf x}_2 ; \omega' )
\right\rangle
=
2\pi \delta( \omega - \omega')
\,
{\cal B}_{ij}( {\bf x}_1, {\bf x}_2 )
\end{equation}
Whenever no confusion is possible, we suppress the frequency dependence 
for simplicity.
At low frequencies, the magnetic noise spectrum tends towards a constant 
anyway.

The spatial Fourier transformed magnetic field~(\ref{eq:Bt-solution})
thus yields, using the current correlation function~(\ref{eq:JJ-correlation}),
\begin{eqnarray}
{\cal B}_{ij}( {\bf x}_1, {\bf x}_2 ) &=&
\frac{ \mu_0^2 S }{ 4 }
\int\!\frac{ {\rm d}^2 K }{ 2 (2\pi)^2 K^5 }
{\rm e}^{ {\rm i} {\bf K} \cdot ({\bf x}_2 - {\bf x}_1) 
- K ( z_1 + z_2 ) }
X_{ij}
\nonumber
\\
&&
\\
X_{ij} &=& 
\left(
- {\rm i} K \epsilon_{ikl} k^*_k + 
({\bf n} \times {\bf K})_i k^*_l
\right)
\delta_{lq}
\nonumber
\\
&& {} \times
\left(
{\rm i} K \epsilon_{jpq} k_p + 
({\bf n} \times {\bf K})_j k_q
\right)
\end{eqnarray}
This tensor can be worked out using the relations
\(
{\bf k} \times {\bf k}^* = 
({\bf K} + {\rm i} K {\bf n})
\times
({\bf K} - {\rm i} K {\bf n})
=
2 {\rm i} K {\bf n} \times {\bf K}
\)
and $|{\bf k}|^2 = 2 K^2$, valid in the quasi-static limit. 
We find
\begin{equation}
X_{ij} = K^2 \left(
 2 K^2 \delta_{ij} - k_i k^*_j 
- 2 ({\bf n} \times {\bf K})_i
({\bf n} \times {\bf K})_j 
\right)
\label{eq:Xij-2}
\end{equation}
where the last term includes also the crossed correlations between the 
two terms in Eq.(\ref{eq:Bt-solution}), leading to the minus sign.
For the planar geometry, we can work out the integral over the azimuthal
angle. For simplicity, we focus on the noise tensor at
the same position ${\bf r} = {\bf x}_1 = {\bf x}_2$. The angular 
average (denoted by double brackets) gives
\begin{eqnarray}
&&
\llangle 
k_i k^*_j 
\rrangle =
\frac12 K^2 \Delta_{ij} 
+ K^2 n_i n_j 
\\
&&
\llangle 
({\bf n} \times {\bf K})_i
({\bf n} \times {\bf K})_j 
\rrangle =
\frac12 K^2 \Delta_{ij} 
\\
\Rightarrow&&
\llangle X_{ij} \rrangle =
K^4 
\left(
\frac12 \Delta_{ij}
+ n_i n_j
\right)
\equiv K^4
s_{ij}
\end{eqnarray}
where $\Delta_{ij} = {\rm diag}( 1, 1, 0 )$ is the in-plane Kronecker
symbol and $s_{ij} = {\rm diag}( \frac12, \frac12, 1 )$ an anisotropic
tensor that was also found in Ref.~\cite{Henkel99c},
using the asymptotic expansion of the fully retarded magnetic noise 
tensor. Thanks to the additional term in~(\ref{eq:Xij-2}),
we thus find the correct magnetic correlation tensor. Without this term,
${\rm diag}(\frac32,\frac32,1)$ would have come out.
Let us check the prefactor of the spectrum. It is given by
\begin{eqnarray}
{\cal B}_{ij}( {\bf r}; \omega ) & = &
\frac{ \mu_0^2 S }{ 4 }
\int_0^\infty\!\frac{ {\rm d} K }{ 4\pi K^4 }
{\rm e}^{- 2 K z }
\llangle X_{ij} \rrangle
\\
\nonumber
&=&
\frac{ \mu_0^2 S }{ 32 \pi z } s_{ij}
= 
\frac{ \mu_0^2 \sigma k_{\rm B} T }{ 16 \pi z } s_{ij}
\end{eqnarray}
This is the result given in Eq.(24) of Ref.~\cite{Henkel99c}, taken 
in the quasi-static limit (distance $z$ small compared to the skin depth).
We thus have shown that when the correct boundary conditions for the 
magnetic field are used, the `incoherent summation' approach is 
equivalent to the more rigorous FD theorem.

\subsection{Impact of finite skin depth}
\label{s:skin-regime}

We now discuss what the previous formulas become when the 
relevant geometrical distances are comparable to the skin depth 
$\delta$. 
The transmitted magnetic field takes the form (see 
Appendix~\ref{a:summation})
\begin{eqnarray}
{\bf B}_{\rm t}[ {\bf K} ] & = & 
\frac{ \mu_0 }{ \kappa + K }
{\bf k}_{t}
\left( 
( {\bf n} \times \hat{\bf K}) \cdot {\bf J}[{\bf K}] \right)
\label{eq:Bt-dynamic}
\end{eqnarray}
Here, $\hat{\bf K}$ is the unit vector along ${\bf K}$. 
This result is transverse as it should because the vacuum wave vector
satisfies ${\bf k}_{t}^2 = 0$. It also coincides with 
Eq.(\ref{eq:Bt-solution}) in the quasi-static limit, as a simple
calculation shows.
Note that, again, ${\bf B}_{\rm t} = {\bf 0}$ if ${\bf j} \Vert {\bf n}$.

The magnetic correlation tensor obtained from Eq.(\ref{eq:Bt-dynamic}) 
can be worked out and reduces to the following simple formula
\begin{equation}
{\cal B}_{ij}( {\bf r}; \omega ) = 
\frac{ \mu_0^2 \sigma k_{\rm B} T }{ 2\pi }
s_{ij}
\int_0^\infty\!\frac{ K^3 {\rm d}K \,
{\rm e}^{ - 2 K z } }{
{\rm Re}\, \kappa \, |\kappa + K|^2 }
.
\label{eq:Bij-full}
\end{equation}
It is interesting to note that the anisotropy tensor $s_{ij}
= {\rm diag}( \frac12, \frac12, 1)$ describes
the magnetic noise through the entire distance range. 
The asymptotic limits for the integral are 
\begin{equation}
\int_0^\infty\!\frac{ K^3 {\rm d}K \,
{\rm e}^{ - 2 K z } }{
{\rm Re}\, \kappa \, |\kappa + K|^2 }
\approx \left\{
\begin{array}{ccl}
\displaystyle
\frac{ 1 }{ 8 z },
&& z \ll \delta
\vspace*{2mm}
\\
\displaystyle
\frac{ 3 \delta^3 }{ 16 z^4 },
&& z \gg \delta.
\end{array}
\right.
\label{eq:int-B-asymptotics}
\end{equation}
The second limit reproduces the result obtained asymptotically 
from the exact solution in the skin-dominated regime. 

Let us analyze this skin-dominated limit in more detail
and consider here the case $|q| = {\cal O}(1/\delta) \gg K$.
We then have $\kappa \approx - {\rm i} q \gg K$ and get to leading order 
\begin{eqnarray}
{\bf B}_{\rm t}[ {\bf K} ] & \approx & 
\frac{ {\rm i} \mu_0 }{ q }
{\bf k}_{t}
( {\bf n} \times \hat{\bf K}) \cdot {\bf J}[{\bf K}] 
\label{eq:Bt-skin}
\end{eqnarray}
Comparing to~(\ref{eq:Bt-dynamic}), we see that the skin effect 
effectively prevents the magnetic field from leaking out of the metal. 
It is sufficient to work to this order to get
the large distance asymptotics 
[Eq.(\ref{eq:int-B-asymptotics}), second line].

\subsection{Discussion} 
\label{s:layer-discussion}

We start with a discussion of the change in the power 
laws~(\ref{eq:int-B-asymptotics}), as one changes the distance
from below the skin depth to much larger values. In the short 
distance regime, the effective volume inside the metal that 
contributes to the magnetic noise, is of the order $z^3$, since 
across the distance $z$, absorption is negligible. Adding incoherently 
magnetic fields with an amplitude $\sim 1/z^2$ for each element 
in this volume, gives the $1/z$ power law for the magnetic noise power. 
At larger distances, damping 
in the metal becomes relevant, and one expects only a surface layer of 
volume $\sim z^2 \delta$ to contribute. This leads to an 
scaling $\delta/z^2$ that is not the one found here. 
In fact, as the skin depth gets shorter, the transmission through the metallic 
surface also becomes more inefficient, as discussed in Sec.\ref{s:skin-regime}.
This leads to a reduced transmitted field. The calculation shows that
in Fourier space, one factor $1/K$ becomes $1/q$ 
[Eqs.(\ref{eq:Bt-solution}, \ref{eq:Bt-skin})] so that the transmitted
field scaling changes like $1/z^2 \mapsto \delta/z^3$. We get the 
$\delta^3/z^4$ behaviour by incoherently summing this up over the near-surface
volume $\sim z^2 \delta$. 

Let us compare to results obtained previously for metallic layers 
with finite thickness $t$. Calculations in this geometry have been 
performed by Varpula and Poutanen \cite{Varpula84}, Sidles and 
co-workers \cite{Sidles00}, and Rekdal and co-workers
\cite{Scheel05a}. At low frequencies where the 
skin depth becomes the largest scale, one gets
\begin{equation}
\delta \gg z, t: \qquad
{\cal B}_{ij}( {\bf r}; \omega ) = 
    \frac{ \mu_0^2 \sigma k_{\rm B} T }{ 16\pi }
    s_{ij}
    \frac{ t }{ z ( z + t ) }.
\end{equation}
This is consistent with the simple rule of removing the vacuum half 
space below 
the layer (replace $1/z$ by $1/z - 1/(z+t)$), ignoring the boundary 
conditions at the lower interface. One can actually show that 
sub-layer material add negligible noise
as long as its
conductivity is much less than that of the metallic layer
(see Refs.~\cite{Scheel05a,Zhang05a}).
In the limit of a thin layer, $t \ll z$, the noise is smaller compared 
to a half-space, because it decays like $t/z^2$. For cylindrical 
wires, the noise reduction is even stronger, see
Refs.~\cite{Scheel04a,Henkel01a}.

At higher frequencies, the skin depth becomes shorter, and different 
regimes emerge depending on the relative magnitude of distance $z$ and 
thickness $t$. We focus in the following on the thin layer limit.
Varpula and Poutanen \cite{Varpula84} give the following empirical 
interpolation formula for a layer thinner than the skin depth
\begin{equation}
t \ll \delta, z: \qquad
{\cal B}_{ij}( {\bf r}; \omega ) \approx
    \frac{ \mu_0^2 \sigma k_{\rm B} T }{ 16\pi }
    s_{ij}
    \frac{ t / z^2 }{ 1 + [4 z t / (\pi^2 \delta^2 ) ]^2  }.
\label{eq:Varpula-interp}
\end{equation}
This gives at large distance $z \gg \delta$ a noise spectrum with 
a scaling $\sim \delta^4/(t z^4)$, similar to the half-space, but with an 
increased amplitude (by a factor of order $\delta / t$). Note that in 
this regime, decreasing the layer thickness just produces the opposite 
effect on the noise. This 
unusual result has been confirmed experimentally in the kHz range 
\cite{Varpula84}. A non-monotonic behaviour with either skin depth 
or conductivity $\sigma = 2 / (\mu_{0} \omega \delta^2)$ 
has also been pointed out by Rekdal and co-workers 
\cite{Scheel04a,Scheel05a}: bad conductors show only weak current fluctuations, 
while good conductors efficiently screen the magnetic 
field.~\footnote{A similar situation occurs for the absorption 
of normally incident plane waves in a thin metallic film.
The maximum absorption occurs for $\delta \sim \sqrt{ \lambda t }$.
See~\cite{Sidles00} for the link to magnetic noise and, e.g.,
\cite{Bauer92} for an instructive discussion.}
Eq.(\ref{eq:Varpula-interp}) indeed yields a noise maximum for 
$\delta \sim \sqrt{ z t }$, consistent with Ref.~\cite{Scheel05a}.

Finally, a layer thicker than the skin depth has been considered in
Refs.~\cite{Scheel05a,Sidles00}, where the following asymptotics is derived
\begin{equation}
{\cal B}_{ij}( {\bf r}; \omega ) \approx
    \frac{ \mu_0^2 \sigma k_{\rm B} T }{ 16\pi }
    s_{ij}\times
    \left\{
    \begin{array}{ll}\displaystyle
	\frac{ \delta^4 }{ 2 t z^4 } & \qquad
	t \ll \delta \ll \sqrt{ z t }
	\\[2ex]
	\displaystyle
	\frac{ 3 \delta^3 }{ 2 z^4 } & \qquad
	\delta \ll \min{( z, t )}
    \end{array}
    \right.
\label{eq:Sidles-asymp}
\end{equation}
Note that the first line differs from Eq.(\ref{eq:Varpula-interp}) by 
a numerical factor -- this may be due to the chosen interpolation.
The second line, consistent with Ref.~\cite{Scheel05a},
shows that for a very short skin depth, there is no 
difference between a metallic layer and a half-space 
[Eq.(\ref{eq:int-B-asymptotics})]. This could have 
been expected given the highly efficient screening.

Let us finally touch upon the case of a superconducting object.
Sidles and co-workers \cite{Sidles00} have argued that it suffices 
to use a 
complex conductivity
and to make the replacement $\sigma \mapsto {\rm Re}[ \sigma( \omega )Ê]$.
For an ideal superconductor and zero temperature, 
London's equation yields
$\sigma( \omega ) 
= - {\rm i} \lambda_{L}^2 / (\mu_{0} \omega )$ with $\lambda_{L}$ the
London penetration depth, and magnetic near field noise is completely 
suppressed. At finite temperature, the superconducting 
phase coexists with a normal phase, and ${\rm Re}[ \sigma( \omega )Ê]$
is finite. In terms of the (frequency-dependent) phase angle $\varphi$ 
in $\sigma = |\sigma| {\rm e}^{-{\rm i}\varphi}$, the following 
interpolation formula is given in Ref.~\cite{Sidles00} for the magnetic 
noise spectrum above a superconducting layer%
\footnote{We have corrected an obvious error in Eq.(6a) of 
Ref.~\cite{Sidles00} and took the classical limit $\hbar\omega \ll 
k_{B}T$.}
\begin{eqnarray}
    {\cal B}_{ij}( {\bf r}; \omega ) &\approx&
	\frac{ \mu_0^2 k_{\rm B} T |\sigma| \cos\varphi }{ 16\pi }
	s_{ij}
	\frac{ 3 \delta^3 t }{ {\cal D} }
\label{eq:Sidles-interp}
\\
{\cal D} &=&
3 \delta^3 z ( z +t ) + 
2\sqrt{2} (1 - {\rm e}^{ - \alpha_{c} })\times
\nonumber
\\
&& {} \times
t z^2 ( z + \delta \sqrt{2})^2 \cos( \pi/4 - \varphi/2)
\\
\alpha_{c} &=&
\frac{ zt + 4 \delta^2 \sin\varphi }{ \sqrt{ 2} \, z \delta 
\cos( \pi/4 - \varphi/2) }
.
\end{eqnarray}
Numerically, it is found that this formula reproduces the results of an exact
calculation to within 2\,{\rm dB}. For a normal conductor ($\varphi = 
0$), it reproduces the asymptotics~(\ref{eq:int-B-asymptotics}, 
\ref{eq:Sidles-asymp}).

Sidles and co-workers have also given corrections for a material with a
weak magnetic susceptibility ($|\mu - \mu_{0}| > 0$) where thermal 
magnetization fluctuations contribute to the magnetic field noise as 
well (with a noise spectrum proportional to ${\rm Im}\,1/\mu$),
see Eqs.(35, 36a) of Ref.~\cite{Sidles00}. Rekdal and co-workers
\cite{Scheel05a} have pointed out that measurements of the magnetic 
susceptibility actually allow to infer the frequency-dependent 
complex conductivity that determines both the skin depth and magnetic
noise properties. For niobium, the skin depth significantly differs from the 
London penetration depth at temperatures below the transition point and 
frequencies around $500\,{\rm kHz}$.

\section{Magnetic Green function}
\label{s:FD-technique}

In this section, we switch to an alternative approach to magnetic 
near field noise that exploits a link to classical dipole radiation.
In fact, the field radiated by a single point-like magnetic moment is 
sufficient to get the magnetic noise spectrum when the 
fluctuation-dissipation theorem~(\ref{eq:FD-magnetic}) is used. 
This is a significant advantage for 
numerical calculations that are needed anyway in more complex 
geometries. In the `incoherent summation' technique, one not only
faces a similar effort to be invested in the computation of the field, 
but the calculation has to be repeated for a large number of
inequivalent sources (all volume elements filled with absorbing
material).  As pointed out by Sidles et al.~\cite{Sidles00}, this 
redundancy can be avoided using the reciprocity theorem: once the
field emitted by a suitable point source is computed, the relevant
quantity is the total power absorbed in the metallic structure.  In
the following, we formulate the equations to solve near arbitrary
metallic structures, with retardation in vacuum being neglected.
Subsection~\ref{s:FD-planar} specializes to a half-space and shows
that the reflected Green tensor is consistent with a fully retarded
calculation.

\subsection{Fluctuation-dissipation theorem}

The fluctuation-dissipation (FD) theorem for the 
magnetic field reads~\cite{Agarwal75a}:
\begin{equation}
    {\cal B}_{ij}( {\bf x}_{1}, {\bf x}_{2}; \omega ) 
    = 2 \hbar f( \hbar \omega / k_{B} T )\,
    {\rm Im}\,{\cal H}_{ij}( {\bf x}_{1}, {\bf x}_{2}; \omega ) 
    \label{eq:FD-magnetic}
\end{equation}
where ${\cal H}$ is the magnetic Green tensor, i.e., the magnetic 
field radiated by an oscillating magnetic point dipole 
${\bf m}$ at ${\bf x}_{2}$,
${B}_{{\rm dip},i}( {\bf x}_{1}; \omega ) = 
\sum_{j}{\cal H}_{ij}( {\bf x}_{1}, {\bf x}_{2}; \omega ) m_{j}$.
For the low-frequency regime relevant here, the temperature 
dependence reduces to $f( \hbar \omega / k_{B} T ) = k_{B} T / \hbar 
\omega$.


Since we are interested in atoms trapped in vacuum above a metallic 
structure, we shall take ${\bf x}_{1} = {\bf x}_{2}$ in 
vacuum.
The magnetic dipole field ${\bf B}_{{\rm dip}}$ then can always be written 
as the sum of the vacuum radiation plus a field scattered or reflected 
from the structure. The vacuum field gives an imaginary part
${\rm Im}\,{\cal H}^{\rm vac}( {\bf x}_{1}, {\bf x}_{1}; \omega )$ that 
reproduces Planck's formula for the blackbody radiation spectrum. We 
shall actually neglect this contribution compared to the one of the 
scattered field. 
For a planar surface, the scattered field can be written as an 
integral over Fresnel reflection coefficients (see, e.g., 
Ref.~\cite{Henkel99c}). We check in the 
following section that the boundary conditions of 
Sec.~\ref{s:jumps} reproduce the Fresnel coefficients, at least in the 
low-frequency limit we focus on here.

We shall use the vector potential ${\bf A}$ and the scalar potential 
$\phi$ in the `generalised Coulomb gauge' $\nabla \cdot \varepsilon 
{\bf A} = 0$. This gives the following wave equations for the domains 
outside and inside the metallic objects whose shape is left arbitrary for 
the moment. Outside the object:
\begin{eqnarray}
\nabla^2 \phi &=& 0
\\
\nabla^2 {\bf A} &=& 
- \mu_{0}\nabla \times {\bf M} -
\frac{i \omega }{c^2} \nabla \phi 
\label{eq:A-outside}
\\
\nabla \cdot {\bf A} &=& 0
\label{eq:div-A-0}
\end{eqnarray}
where ${\bf M}( {\bf x} ) = {\bf m} \, \delta( {\bf x} - {\bf x}_1 )$ 
is the magnetization density for a point dipole. Inside the object:
\begin{eqnarray}
\nabla^2 \phi &=& 0
\\
\nabla^2 {\bf A} + q^2 {\bf A}
&=& \frac{ q^2 }{\omega} \nabla \phi 
\\
\nabla \cdot {\bf A} &=& 0
\end{eqnarray}
We will consistently work in the limit
$|q c / \omega| \sim \lambda / \delta \to \infty$, 
with spatial derivatives being 
comparable to $q$. We thus cover length scales comparable with or smaller
than the skin depth. Combined with the boundary conditions for the 
potentials, these equations allow to determine the field everywhere.
Note that there is no source term in the equations for the scalar 
potential $\phi$, even when the boundary conditions are taken into 
account. Without loss of generality, we therefore put $\phi \equiv 0$ 
in the following.

\subsection{Planar geometry}
\label{s:FD-planar}

In the planar case, we have a simple analytical solution for the
magnetic noise tensor -- a benchmark result that has to be 
reproduced by our theory. Details of the calculation can be found in 
Appendix~\ref{b:FD}. We use the boundary condition 
${\bf n} \cdot \left. {\bf A} \right|_{\rm in} = 0$ characteristic 
for the metal-vacuum interface [Eq.(\ref{eq:normal-E-2})] 
that follows from ${\bf E} = i \omega {\bf A}$. 
Translational symmetry allows to expand the field radiated by the 
magnetic dipole in vacuum in plane waves with wave vector ${\bf K}$ 
parallel to the interface. Focussing on an incident 
plane wave, the calculation yields a reflected magnetic field
with Fourier amplitude
\begin{equation}
    {\bf B}_{r}[ {\bf K} ] = \frac{K - \kappa}{K + \kappa} 
    {\bf k}_{r} 
    \left( {\bf n} \times \hat{\bf K} \right) \cdot
    {\bf A}_{i}[ {\bf K} ]
    ,
    \label{eq:Br-result}
\end{equation}
where ${\bf A}_{i}[ {\bf K} ]$ is given in Eq.(\ref{eq:AiK-above-surface}).
The ratio $(K - \kappa)/(K + \kappa)$ is the same reflection 
coefficient that appears in the (retarded) magnetic Green function,
when the limit $\lambda \to \infty$ is taken,
see, e.g.,~\cite{Henkel99c}. 
In Appendix~\ref{b:FD}, the magnetic Green tensor computed 
from~(\ref{eq:Br-result}) is found to be
\begin{equation}
{\cal B}_{ij}( {\bf r}; \omega ) = 
-
\frac{ \mu_0 k_{\rm B} T }{ 2\pi \omega }
s_{ij}
\int_0^\infty\!\frac{ K^3 {\rm d}K \,
2 \,{\rm Im}\, \kappa\,{\rm e}^{ - 2 K z } }{
 |\kappa + K|^2 }
.
\label{eq:Bij-full-2}
\end{equation}
This expression agrees with Eq.(\ref{eq:Bij-full}) thanks to the 
identity $2\,{\rm Im}\,\kappa\,{\rm Re}\,\kappa = {\rm Im}\,\kappa^2
= - {\rm Im}\,q^2 = - \mu_{0}\sigma\omega$.

\section{Summary and conclusion}

We have discussed in this paper calculations for low-frequency 
magnetic noise fields at sub-wavelength distances to metallic objects. 
The role of the object surfaces has been clarified: they screen some 
field components so that only current elements parallel to the 
surface produce fields outside the object. This occurs even on a 
distance scale where dissipation in the metal is negligible. 
Neglecting this effect leads to errors up to a factor of three for the 
components of the magnetic noise tensor at short distance (smaller 
than the skin depth $\delta$), and completely wrong power laws at 
larger distances ($\gg \delta$). As a consequence, the simple 
incoherent addition of thermal noise fields has to be replaced by a 
more involved calculation, preferably based on the 
fluctuation-dissipation theorem for the field. We have formulated 
an outline of this 
calculation in a generic geometry, spelling out the boundary
conditions that apply in the low-frequency regime characteristic for
miniaturized atom traps. We hope that this opens a way to accurate 
and numerically efficient methods of characterizing magnetic noise 
spectra near complex atom chip structures.

Our theory can be extended to dielectric objects as well. If 
absorption is large and $|\varepsilon| \gg 1$, the same approach 
can be carried over, with the 
skin depth defined by $1/\delta = (\omega/c)\,{\rm Im}\,\sqrt{ 
\varepsilonÊ}$.
For a purely real permittivity, however, one has to include
retardation in the vacuum regions to get a nonzero imaginary part in
the magnetic Green function. When $\varepsilon$ is of order unity,
the boundary conditions for the fields assume, of course, their 
standard form. Numerical calculations are currently under way to 
test the validity of the non-retarded approach.

\bigskip

\noindent
{\footnotesize
I thank Isabelle Bouchoule for stimulating comments and the 
Laboratoire Charles Fabry of the Institut d'Optique for its kind 
hospitality. This work was supported by the European Commission in 
the project ACQP (contract IST-2001-38863) and the network FASTNet
(contract HPRN-CT-2002-00304).}

\appendix

\section{Transmission through a planar interface}
\label{a:summation}

We first compute the magnetic field radiated by a current distribution
${\bf j}( {\bf x} )$ localized inside a homogeneous metal. This is 
given by
\begin{equation}
{\bf B}_{\rm i}( {\bf r} ) = 
\frac{ \mu_0 }{ 4 \pi }
\nabla_{\bf r} \times 
\int\!{\rm d}V( {\bf x} ) \,
\frac{ {\rm e}^{ {\rm i} q |{\bf r} - {\bf x}|} 
}{ | {\bf r} - {\bf x} | }
{\bf j}( {\bf x} ) 
\label{eq:Bi-below-surface}
\end{equation}
where we recall that $q^2 = 2{\rm i} / \delta^2$. 
To solve the transmission problem through a planar interface, it is expedient 
to use the expansion of 
the spherical wave in plane waves (the Weyl angular spectrum)
with the wave vector ${\bf K} = (k_x, k_y, 0)$:
\begin{equation}
\frac{ {\rm e}^{ {\rm i} q |{\bf r} - {\bf x}|} 
}{ 4\pi |{\bf r} - {\bf x} | }
= \frac 12 \int\!\frac{ {\rm d}^2 K }{ (2\pi)^2 \kappa }
{\rm e}^{ {\rm i} {\bf K}\cdot({\bf r} - {\bf x}) - 
\kappa |z - x_3| }
\label{eq:Weyl}
\end{equation}
If we consider a point ${\bf r}$ 
just below the interface, the absolute value in 
Eq.(\ref{eq:Weyl}) can be dropped, and we 
get a Fourier coefficient
\begin{equation}
{\bf B}_{\rm i}[ {\bf K} ] = \frac{ {\rm i} \mu_0 }{ 2 \kappa } 
{\bf k}_{i} \times {\bf J}[ {\bf K} ]
\label{eq:Bi-2}
\end{equation}
where ${\bf k}_{i} = {\bf K} + {\rm i}\kappa {\bf n}$, and ${\bf J}[ {\bf K} ]$
is given by Eq.(\ref{eq:def-J-K}), with $K$ in the exponential replaced by 
$\kappa$. Only in the quasi-static limit, $\kappa \to K$, and we recover 
Eq.(\ref{eq:Bi-1}).

In the same way, we get the normal derivative
\begin{equation}
\frac{ \partial }{ \partial {n} }
{\bf B}_{i}[ {\bf K} ] 
= - \kappa {\bf B}_{i}[ {\bf K} ]
= - \frac{ {\rm i} \mu_0 }{ 2 } 
{\bf k}_{i} \times {\bf J}[ {\bf K} ]
\label{eq:dBi}
\end{equation}
The first equation reflects the property that the magnetic field created by the
current is propagating (in fact, decaying) in the upward direction. 
For the reflected and transmitted fields, similar relations hold:
\begin{equation}
\frac{ \partial }{ \partial {n} }
{\bf B}_{r}[ {\bf K} ] 
= +\kappa {\bf B}_{r}[ {\bf K} ]
, \qquad
\frac{ \partial }{ \partial {n} }
{\bf B}_{t}[ {\bf K} ] 
= -K {\bf B}_{t}[ {\bf K} ]
.
\label{eq:dBrt}
\end{equation}
The jump condition~(\ref{eq:jump-B-2}) thus gives
\begin{equation}
    - K {\bf B}_{t}[ {\bf K} ] + \kappa 
    \left( {\bf B}_{i}[ {\bf K} ] - {\bf B}_{r}[ {\bf K} ] \right)
    =
    \mu_{0} \sigma {\bf n} \times {\bf E}_{t}[ {\bf K} ]
    .
    \label{eq:jump-B-3}
\end{equation}
It turns out that to proceed, we do not actually need to compute the electric 
field at the metallic surface. We combine the continuity of the 
magnetic field perpendicular to the surface with the corresponding 
component of Eq.(\ref{eq:jump-B-3}) and solve for the transmitted 
field component
\begin{equation}
    {\bf n} \cdot {\bf B}_{t}[ {\bf K} ] = \frac{ 2\kappa }{ \kappa + K }
    {\bf n} \cdot {\bf B}_{i}[ {\bf K} ] 
    \label{eq:Bt-2}
\end{equation}
where we recognize one of the Fresnel coefficients. The component 
along ${\bf K}$ follows from ${\rm div}\, {\bf B}_{t} = 
{\rm i} {\bf k}_{t} \cdot {\bf B}_{t}[ {\bf K} ] = 0$.

The last component is along ${\bf n} \times {\bf K}$, and we can use 
the following trick to show that it actually vanishes. 
The Maxwell equations yield ($\alpha = i, r, t$)
\begin{equation}
    {\rm i} ( {\bf n} \times {\bf K} ) \cdot {\bf B}_{\alpha}[ {\bf K} ]
    = 
    {\rm i} {\bf n} \cdot ( {\bf k}_{\alpha} \times 
    {\bf B}_{\alpha}[ {\bf K} ] )
    = 
    \mu_{0} \sigma_{\alpha} 
{\bf n} \cdot {\bf E}_{\alpha}[ {\bf K} ] 
\label{eq:trick-E-B}
\end{equation}
with $\sigma_{i,r} = \sigma$ and $\sigma_{t} \to 0$. We now use the 
boundary condition that the normal electric field
is zero inside the metal [Eq.(\ref{eq:normal-E-2})], 
and get from Eq.(\ref{eq:trick-E-B})
$( {\bf n} \times {\bf K} ) \cdot ( {\bf B}_{i}[ {\bf K} ] 
+ {\bf B}_{r}[ {\bf K} ] ) = 0$. Now, this is a magnetic field component 
tangential to the surface, and therefore
\begin{equation}
    ( {\bf n} \times {\bf K} ) \cdot {\bf B}_{t}[ {\bf K} ] = 0
    .
\end{equation}
Combining the two nonzero components found above, we readily get the 
transmitted field~(\ref{eq:Bt-dynamic}).

\section{Magnetic Green tensor}
\label{b:FD}

The vector potential created by a magnetic point dipole in free space
(solution of Eq.(\ref{eq:A-outside})) is of the form
\begin{equation}
{\bf A}_{i}( {\bf r} ) = 
-
\frac{ \mu_{0} }{ 4 \pi }
\nabla_{\bf r} \times 
\frac{ {\bf m}  }{ | {\bf r} - {\bf x}_{1} | }
\label{eq:Ai-above-surface}
\end{equation}
where ${\bf x}_{1}$ is the dipole position. Above a planar surface, 
we use the  Weyl 
expansion~(\ref{eq:Weyl}) and find the following Fourier coefficient 
for the field incident on the interface:
\begin{equation}
{\bf A}_{i}[ {\bf K} ] = -\frac{ {\rm i} \mu_{0} }{ 2 K } 
{\bf k}_{i} \times {\bf m}\,
{\rm e}^{ - {\rm i} {\bf k}_{i} \cdot {\bf x}_{1} } 
\label{eq:AiK-above-surface}
\end{equation}
with 
an incident wavevector ${\bf k}_{i} = {\bf K} -{\rm i}K{\bf n}$.
The reflected field is characterized by a wavevector 
${\bf k}_{r} = {\bf K}+ {\rm i}K {\bf n}$ and transversal as well, i.e.
${\bf k}_{r} \cdot {\bf A}_{r}[ {\bf K} ] = 0$.

The transmitted field has the wavevector ${\bf k}_{t} = {\bf K}
-{\rm i}\kappa {\bf n}$ with $K^2 - \kappa^2 = q^2$ (${\rm Re}\,\kappa 
> 0$). Here, we have two `transversality conditions':
\begin{equation}
    {\bf k}_{t} \cdot {\bf A}_{t}[ {\bf K} ] = 0,
    \qquad
    {\bf n} \cdot {\bf A}_{t}[ {\bf K} ] = 0,
\end{equation}
where the second one actually comes from the boundary condition for the 
electric field, Eq.(\ref{eq:normal-E-2}).
We conclude that the transmitted field is parallel to the vector
${\bf n} \times {\bf k}_{t} = {\bf n} \times {\bf K}$. 
This vector lies inside the boundary and
is perpendicular to ${\bf K}$.
Since the tangential components of the vector potential are continuous,
the reflected field ${\bf A}_{r}$ has 
to cancel the component of ${\bf A}_{i}$ parallel to
${\bf K}$. This gives a first condition 
for the reflected field:
\begin{equation}
    {\bf K} \cdot \left( {\bf A}_{i}[ {\bf K} ] +  {\bf A}_{r}[ {\bf K} ] 
    \right)
    = 0.
    \label{eq:fix-Ar-parallel}
\end{equation} 

We need a second boundary condition to solve the problem. This is,
of course, the continuity of the magnetic field ${\bf B} = \nabla 
\times {\bf A}$. The magnetic field on the inner side of the interface
is computed to be
\begin{equation}
    \left. {\bf B} \right|_{\rm in} = 
    {\rm i} {\bf k}_{t} \times {\bf A}_{t} = 
    {\rm i} \left( {\bf n} K^2 + {\rm i} \kappa {\bf K} \right) t
\end{equation}
where $t$ is an un-normalized transmission coefficient. 
Computing the normal and ${\bf K}$ component on the `outer side',
we get the linear system
\begin{eqnarray}
K^2 t & \stackrel{^{!}}{=} &
    - {\rm i} {\bf n} \cdot \left. {\bf B} \right|_{\rm out} 
    =  
    \left( {\bf n} \times {\bf K} \right) \cdot 
    \left( {\bf A}_{i} +  {\bf A}_{r} \right)
\\
{\rm i} \kappa K^2 t & \stackrel{^{!}}{=} &
    - {\rm i} {\bf K} \cdot \left. {\bf B} \right|_{\rm out} 
    =  
    {\rm i} K \left( {\bf n} \times {\bf K} \right) \cdot 
    \left( {\bf A}_{i} -  {\bf A}_{r} \right)
\end{eqnarray}
whose solution involve the standard Fresnel reflection and
transmission coefficients:
\begin{eqnarray}
   t & = & \frac{2}{K (K + \kappa)} 
   \left( {\bf n} \times {\bf K} \right) \cdot 
   {\bf A}_{i},
   \\
   \left( {\bf n} \times {\bf K} \right) \cdot 
   {\bf A}_{r}
   & = & \frac{K - \kappa}{K + \kappa} 
   \left( {\bf n} \times {\bf K} \right) \cdot 
   {\bf A}_{i}.
\end{eqnarray}
This yields the following expression for the reflected field
\begin{equation}
    {\bf A}_{r} =
    - \frac{ {\bf k}_{r} }{ K }\left( \hat{\bf K} \cdot {\bf A}_{i} \right)
    + \frac{K - \kappa}{K + \kappa} 
    \left( {\bf n} \times \hat{\bf K} \right) 
    \left( {\bf n} \times \hat{\bf K} \right) \cdot
    {\bf A}_{i}
    \label{eq:Ar-result}
\end{equation}
whose first (`longitudinal') term ensures condition 
(\ref{eq:fix-Ar-parallel}) while still being `transversal' (this is 
due to the fact that ${\bf k}_{r}^2 = 0$ in the
limit $\lambda \to \infty$). The corresponding magnetic field 
is determined by the second term only and one finds Eq.(\ref{eq:Br-result}).

From Eq.(\ref{eq:Ar-result}), we identify the following magnetic Green tensor 
\begin{eqnarray}
    {\cal H}( {\bf r}, {\bf r} ) &=& 
    {\cal H}^{\rm (vac)}( {\bf r}, {\bf r} ) +
    {\cal H}^{\rm (ref)}( {\bf r}, {\bf r} )
    \\
    {\cal H}^{\rm (ref)}( {\bf r}, {\bf r} ) &=&
    \int\!\frac{ d^2K }{ (2\pi)^2 }
    \frac{  \mu_{0} \, {\rm e}^{-2K z} }{ 2 K }
    \frac{K - \kappa}{K + \kappa} 
    {\bf k}_{r} \otimes {\bf k}_{i}
\end{eqnarray}
using the identity $( {\bf n} \times \hat{\bf K} ) \times {\bf k}_{i} = 
{\rm i} {\bf k}_{i}$.
The integration over the azimuthal angle amounts to angular averaging:
$\llangle {\bf k}_{r} \otimes {\bf k}_{i} \rrangle = K^2 s_{ij}$.
For the fluctuation-dissipation theorem, we need the imaginary part
of the reflection coefficient
\begin{equation}
    {\rm Im}\,\frac{ K - \kappa}{K + \kappa} = 
    \frac{ - 2 K \, {\rm Im}\,\kappa }{ |K + \kappa|^2 }
.
\end{equation}
The vacuum Green tensor is purely real in the static limit $\lambda \to 
\infty$ we focus on here, and does not contribute to the magnetic 
noise spectrum~(\ref{eq:FD-magnetic}).
Putting everything together, we get the Green tensor~(\ref{eq:Bij-full-2}).

%

\end{document}